\documentclass[twocolumn,amsmath,amssymb,citeautoscript,prb]{revtex4}

\usepackage{graphicx}
\usepackage{MnSymbol}
\usepackage{color}
\usepackage{hyperref}

\begin{document}

\title{Metallic behaviour in SOI quantum wells with strong intervalley scattering}

\author{V. T. Renard$^1$\footnote{Correspondance to : vincent.renard@cea.fr}}
\author{I. Duchemin$^2$}
\author{Y. Niida$^3$}
\author{A. Fujiwara$^4$}
\author{Y. Hirayama$^3$}
\author{K. Takashina$^5$}
\affiliation{$^1$SPSMS, UMR-E 9001, CEA-INAC/UJF-Grenoble 1, INAC, Grenoble F-38054, France\\
$^2$SP2M, UMR-E 9001, CEA-INAC/UJF-Grenoble 1, INAC, Grenoble F-38054, France\\
$^3$Graduate School of Science, Tohoku University, Sendai-shi, Miyagi 980-8578 Japan\\
$^4$NTT Basic Research Laboratories, NTT Corporation, Atsugi-shi, Kanagawa 243-0198, Japan\\
$^5$Department of Physics, University of Bath, Bath BA2 7AY, UK
}

\date{\today}
\maketitle

\textbf{The fundamental properties of valleys are recently attracting growing attention due to electrons in new and topical materials possessing this degree-of-freedom and recent proposals for valleytronics devices. In silicon MOSFETs, the interest has a longer history since the valley degree of freedom had been identified as a key parameter in the observation of the controversial ``metallic behaviour'' in two dimensions. However, while it has been recently demonstrated that lifting valley degeneracy can destroy the metallic behaviour, little is known about the role of intervalley scattering. Here, we show that the metallic behaviour can be observed in the presence of strong intervalley scattering in silicon on insulator (SOI) quantum wells. Analysis of the conductivity in terms of quantum corrections reveals that interactions are much stronger in SOI than in conventional MOSFETs, leading to the metallic behaviour despite the strong intervalley scattering.}

The prospect of manipulating the valley degree of freedom in materials like AlAs,\cite{Shayegan2004} silicon\cite{Takashina2004,Takashina2006,Culcer2012} graphene,\cite{Beenakker2007} MoS$_2$,\cite{Behnia2012,Zeng2012,Mak2012,Cao2012} or silicene\cite{Tabert2013} is stimulating considerable interest as it could lead to a new paradigm in technology exploiting valleys in addition to charge and spin: valleytronics. From the fundamental point of view, valley-physics provides new insight into fundamental questions such as whether or not electrons confined to two dimensions conduct electricity when the temperature approaches absolute zero and which remains a subject of intense research and controversy despite decades of research. In the 1970's, the scaling theory of conductance established that at $T=0$~K, a non-interacting two-dimensional electron gas (2DEG) would be localised by any amount of disorder at sufficiently large scales.\cite{Abrahams1979} This prediction was verified experimentally rapidly \cite{Dolan1979} and shaped our understanding of the two-dimensional electron gas for the next two decades despite theoretical refinements taking into account electron-electron interaction predicting the possibility of delocalisation.\cite{Altshuler1980,Finkelstein1983,Castellani1984,Finkelstein1984,Gold1986,Zala} Valley degeneracy increases the effect of interactions and indeed, the 2DEG in silicon MOSFETs which possesses this degeneracy, was the system in which strong metallic behaviour was first observed.\cite{Cham1980,Dorozhkin1984,Kravchenko1994} The precise understanding of real systems is however complicated because valley degeneracy is accompanied by intervalley scattering and possibly valley splitting. \cite{Punnoose2001,Vitkalov2003,Klimov2008,Burmistrov2008,Punnoose2010,Punnoose2010b} Both these mechanisms are predicted to weaken the delocalising effect of interactions by reducing the number of active interaction channels.\cite{Punnoose2001} While it is now well established experimentally that lifting valley degeneracy can destroys the metallic behaviour,\cite{Shayegan2007,Takashina2011} less is known about the effect of intervalley scattering. In order to fully understand the role played by intervalley scattering, it is important to compare the behaviour of systems ranging from those with very weak intervalley scattering to strong intervalley scattering. However, previous studies in which the role of intervalley scattering on interactions has been analysed had been limited to systems where this scattering is only weak or moderate so that its effect on the metallic behavour is either absent or restricted to the lowest temperatures.\cite{Vitkalov2003,Klimov2008,Punnoose2010} Here, we investigate two-dimensionnal electron gases confined in a SOI quantum well \cite{Takashina2004,Takashina2006,Takashina2011} (See Methods) where intervalley scattering is an order of magnitude stronger than previously studied so that it modifies the interaction at temperatures up to about 10 K. In addition, compared to Si MOSFETs, interactions are particularly strong in SOI quantum wells because of the reduced dielectric constant $\epsilon$ of the surrounding SiO$_2$. For example, the Wigner-Seitz parameter $r_s \propto 1/\epsilon n^{1/2}$ at 
the density $n=3.9\times10^{15}$ m$^{-2}$ is $r_s=8.2$ compared to $r_s=4.2$ in MOSFETs (in Si MOSFETs, $\epsilon=(\epsilon_{Si}+\epsilon_{SiO_2})/2$=7.7 while in SOI quantum wells $\epsilon\approx \epsilon_{SiO_2}$=3.9). This allows studying a regime where both interactions and temperatures are large while the system remaining degenerate. 

\begin{figure*}[t]
\center
\includegraphics[width=1.5\columnwidth]{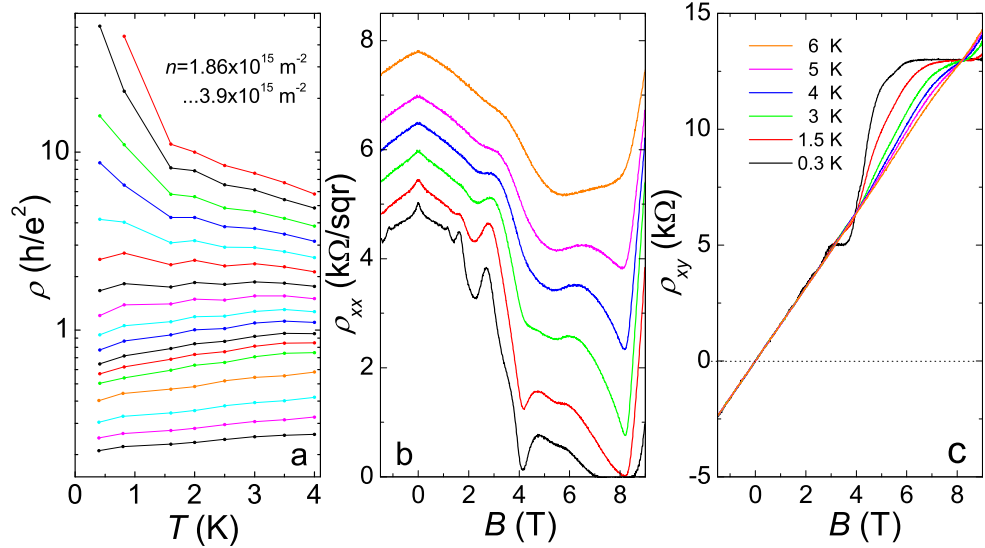}
\caption{Magnetotransport in SOI quantum well. (a) Temperature dependence of the longitudinal resistivity at various densities 
($n= $1.87, 1.95, 2, 2.1, 2.17, 2.25, 2.32, 2.4, 2.47, 2.55, 2.62, 2.7, 2.77,  3, 3.3, 3.6, 3.9$\times10^{15}$ m$^{-2}$ from top to bottom). 
Longitudinal (b) and Hall (c) resistivity of the sample at $n=3.9\times10^{15}~m^{-2}$ and various temperatures (Same temperatures in the 
two panels).\label{Data}}
\end{figure*}

\section*{Results}

The longitudinal resistivity of our sample as function of temperature is displayed in Fig.~1a for different electron densities. An apparent metal-insulator transition is observed 
at $n_c\approx2.3\times10^{15}$ m$^{-2}$ where the resistivity is about $2\times h/e^2$. 
Deep in the metallic regime, the region of interest for the present study, the conductivity at $B=0$ T increases linearly when temperature decreases (Figure~2). 
At $n=3.9\times10^{15}$ m$^{-2}$, where the mobility at $T=0.35$ K is 0.3 m$^2$/Vs, conductivity increases by 60 \% when temperature is decreased from 6 K to 0.35 K. At low fields and low temperatures, 
we attribute the sharp negative magneto-resistance seen in Fig.~1b to weak-localisation.\cite{Hikami}


In the Fermi Liquid theory, weak localisation and electron-electron interactions appear as additive corrections to the classical 
Drude conductivity $\sigma_D$ :\cite{Zala}

\begin{equation}
\sigma(T,B=0)= \sigma_D+\Delta\sigma_{\mathrm{WL}}(T)+\Delta\sigma_{\mathrm{ee}}^{\mathrm{Diff}}(T)+\Delta\sigma_{\mathrm{ee}}^{\mathrm{Ball}}(T)
\label{Conductivity}
\end{equation}

where, $\Delta\sigma_{\mathrm{WL}}$ is the weak localisation (WL) correction, $\Delta\sigma_{\mathrm{ee}}^{\mathrm{Diff}}(T)$ is the electron interaction correction 
in the diffusive regime ($k_BT\tau/\hbar<1$, where $k_B$ and $\hbar$ are Boltzmann and reduced Planck constants and $\tau$ is the Drude scattering time) and 
$\Delta\sigma_{\mathrm{ee}}^{\mathrm{Ball}}(T)$ is the electron interaction correction in the ballistic limit ($k_BT\tau/\hbar>1$). Historically, the diffusive and ballistic interaction corrections had been thought to be governed by a different physics.\cite{Altshuler1980,Gold1986} By studying intermediate temperatures ($k_BT\tau/\hbar\sim1$) where both terms contribute, Zala, Narozhny and Aleiner (ZNA) established that they are governed by the same effect: the scattering off Friedel oscillations, which extend longer or shorter distances compared to the mean free path depending on the regime of interaction.\cite{Zala} Interestingly, the ballistic contribution is predicted to be linear in temperature and might therefore explain the linear temperature dependence observed in Fig.~2. In units of the quantum conductance $e^2/h$, this correction is expressed as :\cite{Zala}


\begin{equation}
\Delta\sigma_{\mathrm{ee}}^{\mathrm{Ball}}(T)= \frac{2k_BT\tau}{\hbar}\times\left( 1+\frac{(4n_v^2-1)F_{0}^{\sigma}}{1+F_0^{\sigma}}\right)
\label{ZNABall}
\end{equation}

where $k_B$ and $h$ are Boltzmann and Plack constants, $n_v$ is the valley degeneracy and $\tau$ is obtained from the Drude conductivity $\sigma_D=ne^2\tau/m^*$ ($e$ is the electron charge and $m^*$ their effective mass). We estimate $\tau$ to be $3.8\times10^{-13}$ s by extrapolating the conductivity to $T=0$~K as suggested in  Ref.~\onlinecite{Zala}. The interaction parameter $F_0^\sigma$ is usually determined from the slope of the conductivity as function of temperature, assuming $(4n_v^2-1)=15$ due to the double valley degeneracy in (001) silicon quantum wells.\cite{Punnoose2001} However, intervalley scattering mixes the valleys in real samples and reduces the number of diffusion channels from 15 to 3 for temperatures lower than $T_\bot=\hbar/k_B \tau_\bot$.\cite{Punnoose2001} Here $\tau_\bot$ is the intervalley scattering time. It is therefore important to determine in which regime of intervalley scattering the sample is in.

Intervalley scattering time can be extracted from the weak localisation magneto-conductivity (MC).\cite{Altshuler1981,Kuntsevich2007} At low magnetic field and for strong intervalley scattering, the MC is that of a single valley system since the valleys are mixed at the time scale of the phase coherence time $\tau_\varphi$. As the magnetic field is increased, the typical length of the trajectories contributing to the WL correction decreases so that the valley mixing along these trajectories decreases. It is suppressed for $B>>B_v\approx \hbar/2eD\tau_\bot$ (Here $D=1/2v_F\tau$ is the diffusion coefficient where $v_F$ is the Fermi velocity). As a result, the system approaches the MC of a system with no intervalley scattering (i.e. two times that of a single valley system). We used a recent theory interpolating between these regimes \cite{Kuntsevich2007} to fit the experimental data. The two adjustable parameters were $\tau_\bot$ and $\tau_\varphi$ (See supplementary information for the code used).

The result of this procedure is displayed in Fig.~3. The theory describes perfectly the experimental data. We checked that the phase coherence time obtained by fitting the MC with the Hikami formula \cite{Hikami} is consistent with that obtained with the equations of Ref~\onlinecite{Kuntsevich2007} (see Fig.~3). The Hikami formula is known to provide good estimates of $\tau_\varphi$ even in systems with intervalley scattering if sufficiently low field is used for the fitting.\cite{Kuntsevich2007} The phase breaking time follows the $T^{-1}$ temperature dependence expected for electron-electron interactions.\cite{Altshuler1982,Narozhny2002} The intervalley scattering time is found to be smaller than the phase coherence time in the entire temperature range 
confirming that intervalley scattering is strong in our sample. In addition, $\tau_\bot$ is temperature independent suggesting that it is governed by static disorder.\cite{Kuntsevich2007}

\begin{figure}
\center
\includegraphics[width=0.9\columnwidth]{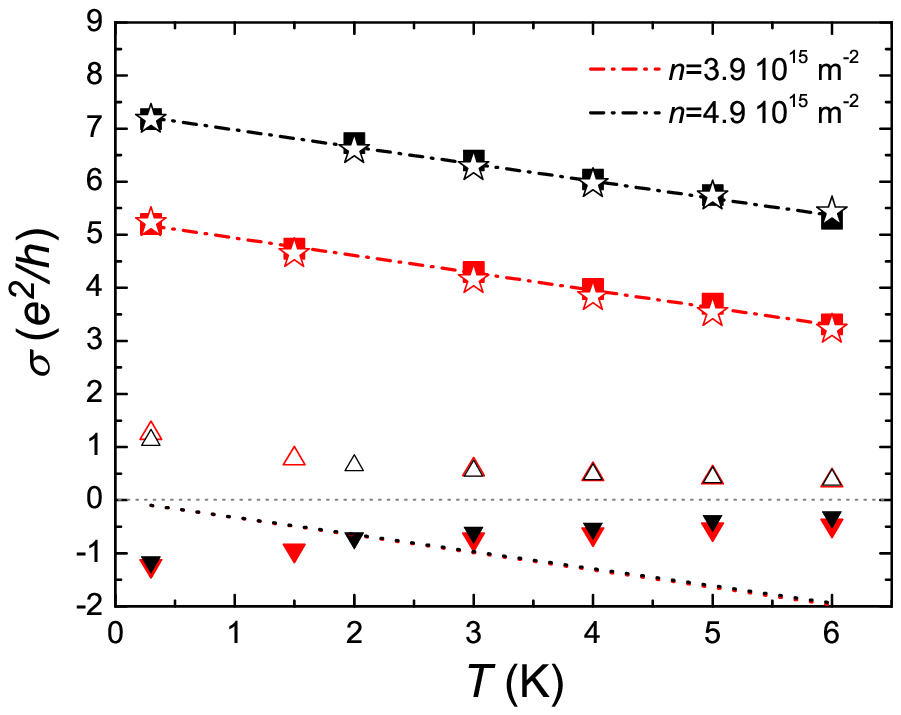}
\caption{Temperature dependence of conductivity. Measured conductivity ($\blacksquare$), fit of $\sigma_D+\sigma_\mathrm{Ball}$ to these data (\textbf{- $\mathord{\cdot}$ -}), Ballistic contribution only ($\mathord{\cdot} \mathord{\cdot} \mathord{\cdot}$),  diffusive contribution ($\triangle$), WL contribution ($\blacktriangledown$) and the sum of Drude, WL, diffusive and ballistic contributions ($\medstar$). Red symbols correspond 
to $n=3.9\times10^{15}$ m$^{-2}$ and the black symbols to $n=4.9\times10^{15}$ m$^{-2}$. The crossover from diffusive to ballistic regime of interaction is illustrated by the ballistic contribution vanishing at low T while the diffusive one vanishes at high T.}
\label{SigmaSummary}
\end{figure} 

Let us now determine the number of active interaction channels in our sample. From the measured value of $\tau_\bot$ we obtain $T_\bot\approx 7.5$ K. This means that our 2DEG behaves like a single valley system from the point of view of interaction corrections to conductivity in the entire range of temperatures. In other words, the factor $(4n_v^2-1)$ should be replaced by 3 in Eq.~\ref{ZNABall}. By fitting the experimental conductivity to $\sigma_D+\Delta\sigma_{\mathrm{ee}}^{\mathrm{Ball}}(T)$ we obtain $F_0^\sigma=-0.59$ and verify that the ballistic contribution of electron-electron interaction correction to conductivity describes our data very well (See Fig.~2). The value of $F_0^\sigma$ seems very large compared to that obtained previously in Si MOSFETs for comparable electron density. However, we recall that in SOI, interactions are enhanced so that from the point of view of interactions, our system behaves like a usual MOSFETs of much lower electron density. At comparable value of $r_s$, $F_0^\sigma$ is also of the order of -0.6 in usual Si MOSFETs (See Fig.~5b in Ref.~\onlinecite{Klimov2008}). This confirms the consistency of the results obtained in the two systems. Here, we should stress that it is very surprising that the ballistic interaction correction can describe alone the conductivity. Indeed, given the measured value of $\tau$, we have $k_BT\tau/\hbar\lesssim1$ and therefore the system is entering the diffusive regime. In addition, as we have already observed, WL produces a large MC and should therefore also contribute significantly to the temperature dependence of conductivity at $B=0$ T. This paradox of the coexistence of weak localisation magneto-resistance and the absence of its temperature dependence at $B=0$ T is evident in previously reported data \cite{Brunthaller2001} but has remained largely overlooked despite an interesting interpretation was suggested in Ref.~\onlinecite{Coleridge2002}. Following and extending the analysis of Ref.~\onlinecite{Coleridge2002}, we now show how the paradox is resolved in our case.

At $k_BT\tau/\hbar\lesssim1$ all contributions to Eq.~\ref{Conductivity} have to be taken into account.\cite{Renard2005,Minkov2006}
The  contribution of weak localisation is readily known once the WL magneto-conductivity is analysed. It reads :\cite{Abrahams1979}  

\begin{equation}
 \Delta\sigma_{\mathrm{WL}}(T)=-\frac{\alpha n_v}{\pi}ln \left( \frac{\tau_{\varphi}}{\tau}\right )
\label{EqTauphi}
\end{equation}

where $\alpha=0.5$ in presence of strong intervalley scattering. This contribution is plotted as filled triangles in Fig.~2. As expected, 
weak localisation does indeed make a large negative contribution, especially at low temperatures. This contribution has to be cancelled by the diffusive interaction 
correction which is predicted to be\cite{Altshuler1980,Zala}:

\begin{figure}
\includegraphics[width=0.9\columnwidth]{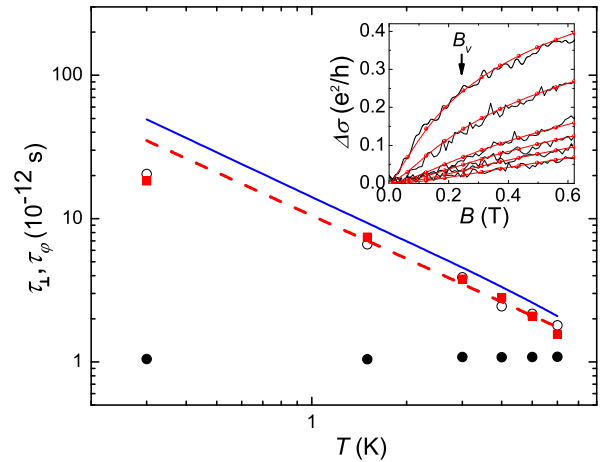}
\caption{Phase coherence and intervalley scattering time. Phase coherence (\textcolor{red}{$\blacksquare$}) and intervalley ($\bullet$) scattering time obtained by fitting with equations of
 Ref.~\onlinecite{Kuntsevich2007}. The black circles ($\circ$) are obtained with Hikami formula for $B<0.15$ T. The dotted red line represents the $T^{-1}$ dependence and 
the blue solid line represents the prediction of ZNA theory with $\mathcal{F}_0^\sigma$=-0.67.  The insert shows the experimental magneto-conductivity (black curves) and result of the fit (red curves) for different temperatures (T=0.3 K, 1.5 K, 3 K, 4 K, 5 K, 6 K from top to bottom.).
The arrow indicates the crossover field $B_v\approx0.25$ T (See the text).
\label{WL}}
\end{figure}

\begin{align}
\begin{split}
\Delta\sigma_{\mathrm{ee}}^{\mathrm{Diff}}(T)&=\frac{1}{\pi}\left( 1+(4n_v^2-1)\left( 1-\frac{ln(1+\mathcal{F}_0^{\sigma})}{\mathcal{F}_0^{\sigma}}\right)\right) \\
&\times ln\left (\frac{k_BT\tau}{\hbar}\right )
\end{split}
\label{EqDiff}
\end{align}
Here again, the factor $4n_v^2-1$ has to be replaced by 3 in our case of strong intervalley scattering. We have used a different symbol for the interaction parameter 
$\mathcal{F}_0^\sigma$ to stress that there is no reason for it to be the same in the diffusive and ballistic regimes \cite{Zala,Renard2005,Minkov2006,Lu2011}. This contribution 
is plotted as open triangles in Figure~2 for $\mathcal{F}_0^\sigma=-0.67$ while the sum of all three contributions is plotted as opened stars. As 
can be seen, the localising effect of WL can indeed be counterbalanced by the delocalising effect of diffusive electron-electron interactions. This is an important 
result of our study which explains the apparent paradox of a ballistic-like temperature dependence of the conductivity while the system is entering the diffusive 
regime of interactions. 
In addition to the conductivity at $B=0$ T, ZNA theory makes a prediction for the temperature dependence of the Hall coefficient \cite{Zala2001c}. Figure~1c shows that in our case the Hall coefficient is surprisingly temperature independent. This observation is in fact very well described by the theory which predicts the absence of temperature dependence of the Hall coefficient for $F_0^\sigma\approx-0.6$ (see Fig.~2 and its caption in Ref.~\onlinecite{Zala2001c}). The theory also provides a prediction for the phase breaking time which can be obtained recursively as described in Ref.~\onlinecite{Narozhny2002}. The solid blue line in Fig.~3 represents this prediction and shows that the theory agrees well with our data. All transport coefficients and the phase coherence time are therefore described consistently and we conclude that the theory captures the observed phenomenology. 

\section*{Discussion}

The cancellation of two quantum effects is a surprising observation and it is interesting to consider to what extent our experimental conditions are special in enabling 
us to observe the cancellation of weak localisation and diffusive interaction correction. Other authors have already reported \cite{Brunthaller2001} on the co-existence of linear temperature dependence of conductivity at zero field and the weak localisation magneto-resistance (i.e. the evidence of the cancellation). We have performed a similar analysis on the data from sample Si15 published in Ref.~\onlinecite{Brunthaller2001}. The result is displayed in Fig.~4. Although intervalley scattering is weak in this sample (we find $T_\perp\approx 80$ mK) and therefore all 15 multiplets contribute to the transport, we observe the cancellation. As expected from the lower interaction in Si MOSFETs, the interaction parameter obtained in Si15 ($F_0^\sigma=-0.30$ and $\mathcal{F}_0^\sigma=-0.31$) is smaller than in our sample for a comparable electron density. The values of interaction parameter are comparable to those obtained previously for similar values of $r_s$.\cite{Klimov2008} In addition, a similar cancellation has also been reported in two-dimensional hole gases in SiGe, a single valley system.\cite{Coleridge2002} In SiGe, the cancellation is found for $\mathcal{F}_0^\sigma\approx -0.7$, comparable to our observation. This confirms the consistency of our analysis since the cancellation should occur for $2+3\left(1-\frac{ln(1+\mathcal{F}_0^{\sigma})}{\mathcal{F}_0^{\sigma}}\right)=0$ in both systems which are single valley from the point of view of quantum corrections. The cancellation is therefore observed in a wide variety of samples with different strength of intervalley scattering and host material and we conclude that although it is not a universal behaviour (see Ref.~\onlinecite{Lu2011} for an illustrative counter example), it is also not an extremely rare effect.
Let us now consider the density dependence of the cancellation. We found that the cancellation is still observed at $n=4.9\times 10^{15}$ cm$^{-2}$ 
(Fig.~2). We interpret this as resulting from the weak density dependence of the interaction parameter. Since $\mathcal{F}_0^\sigma$ weakly 
depends on the density (We found $\tau_\bot=10^{-12}$ s, $F_0^\sigma=-0.57$ and $\mathcal{F}_0^\sigma=-0.64$ at $n=4.9\times 10^{15}$ cm$^{-2}$), 
 $\alpha n_v+1+(4n_v^2-1)\left(1-\frac{ln(1+\mathcal{F}_0^{\sigma})}{\mathcal{F}_0^{\sigma}}\right)\approx0$ over a wide range of densities.

Before concluding we note that in the theoretical expressions of the ZNA theory, equation \ref{ZNABall} contains two dimensionless transition functions ($f(T)$ and $t(T,F_0^\sigma$)) to refine the description of the crossover between diffusive and ballistic limits.\cite{Zala}  In the analysis presented, we have followed the prescription of ZNA which states that ``they can be neglected for all practical purposes",\cite{Zala} as also followed successfully by other experimental studies\cite{Coleridge2002,Minkov2006} and this has enabled us to capture the essential physics of the system. However, when these functions are included, we find that we are not able to match the theory to our data. At present, the extent to which neglecting these functions is justified theoretically remains unclear and an analysis of sufficient detail required for a meaningful treatment is beyond the scope of the present work.

\begin{figure}
\includegraphics[width=0.9\columnwidth]{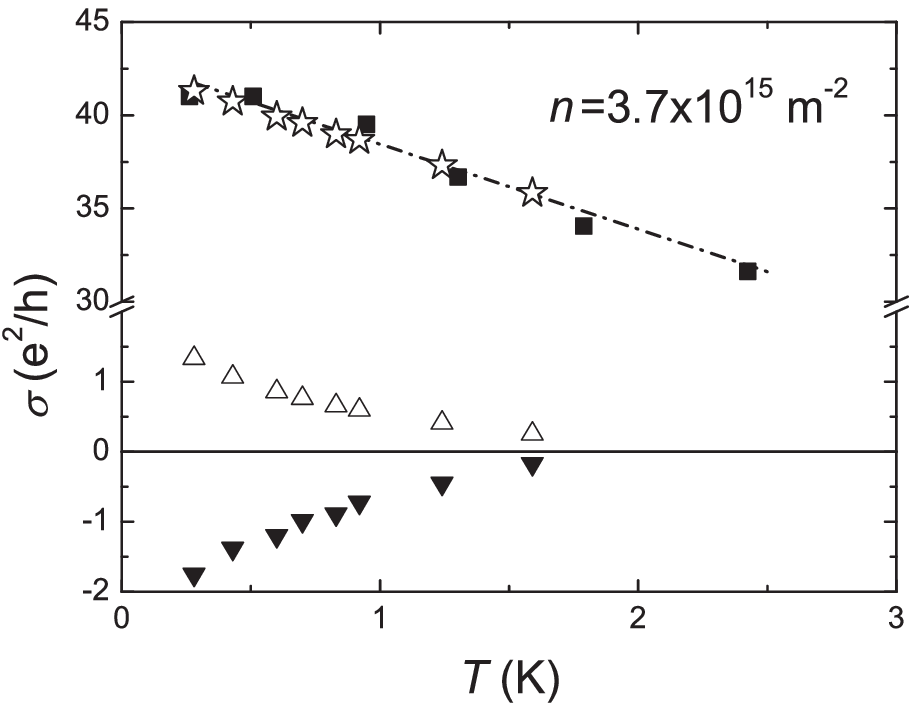}
\caption{Conductivity of the sample Si15 from Ref.~\onlinecite{Brunthaller2001}. The data  ($\blacksquare$) were analysed in the same way as those of Fig.~2.
Fit of $\sigma_D+\sigma_\mathrm{Ball}$ to these data (\textbf{$-\cdot-$}), diffusive contribution ($\triangle$), WL contribution ($\blacktriangledown$) and the sum of Drude, WL, 
diffusive and ballistic contributions ($\medstar$). Despite all interaction channels contribute (see the text), the cancellation of WL and diffusive interaction corrections is observed with $\alpha=1$,
 $F_0^\sigma=-0.3$ and $\mathcal{F}_0^\sigma=-0.31$. Here $r_s\approx 4.3$ compared to 8.2 in our SOI sample. 
\label{Brunthaller_Cancellation}}
\end{figure}

In conclusion, the metallic temperature dependence observed in our sample is quantitatively explained by the leading terms in ZNA theory if strong intervalley scattering is taken into account.
This strong intervalley scattering can explain the modest $T$-dependence in our sample compared to samples with the strongest ``metallicity'' where intervalley
scattering was estimated to be weak \cite{Klimov2008,Punnoose2010} because intervalley scattering reduces the number of diffusion channels contributing to the transport. 
Therefore, instead of their ``quality'', the empirical criterion used until now, our result confirms the suggestion made in Ref.~\onlinecite{Punnoose2001} that intervalley scattering time is the 
quantitative criterion that distinguishes Si samples with strong or moderate metallic temperature dependence. As a consequence we expect that new aspects of the metal insulator transition in 2D could be unveiled in SOI quantum wells with low intervalley scattering where interaction is particularly strong and all interaction channels contribute to the transport. 

\section*{Methods}
The samples consist of a SiO$_2$/Si(100)/SiO$_2$ quantum well of nominally 10 nm thick silicon with a peak mobility $\mu_{peak}=1.0$ m$^2$/Vs.  Hall bar shaped 
samples were measured in a $^3$He cryostat by a standard four-terminal lock-in technique with current below 2 nA to avoid electron heating. 
The fabrication of the samples is described in Ref.~\onlinecite{Takashina2004}. The Silicon on Insulator configuration excludes the possibility of 
any parallel conduction. This is confirmed by the well quantized values of the Hall resistance and good zeros of the longitudinal resistance in the quantum Hall regime.  
Similar results were obtained on different samples from the same wafer and we present the data from only one of them. The data described here were measured at a 
back gate voltage of 2V at which the metallic behavour is maximum (See Ref.~\onlinecite{Takashina2011}). At this gate voltage, 
valley splitting is small \cite{Takashina2006} and was therefore neglected.

 \bibliographystyle{apsrev4-1}

\section*{Acknowledgements}

We are grateful to I. Gornyi for illuminating discussions and to I. Burmistrov, A. Kuntsevich and B. Narozhny for stimulating comments. 
KT is supported by the EPSRC of the UK (EP/I017860/1). YH acknowledges support from JST-ERATO programs. AF acknowledges support from NEXT program (GR103) of JSPS.

\section*{Author contribution}
VTR and KT designed the experiments. AF fabricated the samples. ID wrote the code for the analysis of the WL. VTR performed the experiments and analysis which were discussed by all the authors. VTR wrote the manuscript 
and all co-authors commented on it. KT coordinated the collaboration.

\section*{Additionnal information}

The authors declare no competing financial interest.

\end{document}